\documentclass[runningheads]{llncs}
\usepackage[T1]{fontenc}
\usepackage{graphicx}
\usepackage{amsmath,amssymb}
\usepackage{booktabs}
\usepackage{subcaption}
\usepackage{tikz}
\usepackage{tabularx}
\usepackage{url}
\usepackage{rotating}

\begin{document}

\begin{center}
    \textit{This is the preprint version of a paper accepted for presentation at the 9th International Conference on Communications and Future Internet (ICCFI 2025), Okinawa, Japan, July 11–13, 2025.}
\end{center}

\title{Dual RIS-Assisted Monostatic L-Band Radar Target Detection in NLoS Scenarios}

\author{Salman Liaquat\inst{1} \and Ijaz Haider Naqvi\inst{2} \and Nor Muzlifah Mahyuddin\inst{1}\thanks{Corresponding author: eemnmuzlifah@usm.my}}

\authorrunning{S. Liaquat et al.}

\institute{
School of Electrical and Electronic Engineering, Universiti Sains Malaysia, 14300 Penang, Malaysia 
\and
Department of Electrical Engineering, Lahore University of Management Sciences, Lahore 54792, Pakistan
}

\maketitle

\begin{abstract}
The use of a single Reconfigurable Intelligent Surface (RIS) to boost the signal-to-noise ratio (SNR) at the radar offers significant improvement in detecting targets, especially in non-line-of-sight (NLoS) scenarios. However, there are scenarios where no path exists between the radar and the target, even with a single RIS-assisted radar, due to other present obstacles. This paper derives an expression for SNR in target detection scenarios where dual RISs assist a monostatic radar in NLoS situations. We calculate the power received at the radar through a dual RIS configuration. We show that the SNR performance of RIS-assisted radars can improve with known locations of the radar and RISs. 
Our results demonstrate that the required accuracy in target localization can be achieved by controlling the number of RISs, the number of unit cells in each RIS, and properly selecting the locations of RISs to cover the desired region.
The performance of dual RIS-assisted radar systems can surpass that of single RIS-assisted radar systems under favourable alignment and sufficiently large RIS sizes.


\end{abstract}
\section{Introduction}
Reconfigurable Intelligent Surfaces (RISs) have garnered considerable interest in wireless communications, establishing the groundwork for smart radio environments, particularly for next-generation communication (beyond 6G) and sensing systems \cite{ElMossallamy_reconfigurable_2020,tang2022path,liu_reconfigurable_intelligent_2021}. 
These nearly passive, planar structures are composed of a large number of sub-wavelength unit cells that exhibit tunable electromagnetic characteristics, allowing them to manipulate existing waves without emitting any power of their own \cite{bjornson_reconfigurable_2020,basar2019wireless}. By altering the amplitude, phase, frequency, and polarization of impinging signals, RISs can effectively modify the communication channel \cite{zhang2020optically}. Leveraging the cutting-edge capabilities of RISs, researchers have recently focused on revolutionizing sensing systems and fostering smart radio environments \cite{wu_towards_2020,di_renzo_2020_smart_radio}. These surfaces may be installed outdoors on building facades or indoors on the ceiling or walls \cite{buzzi_radar_2021}.
RIS is an active area of research in communications, with fundamental roots in the electromagnetic field. A common feature on all RISs is the presence of several discrete elements with controllable attributes. A significant study has been conducted on how to produce such surfaces and regulate their characteristics, as well as on implementation methods using different materials for different frequencies and application cases \cite{di_renzo_communication_2022}.

Recently, using RIS in radars has garnered interest from researchers \cite{lu_target_2021,buzzi_foundations_2022,haobo_metardar_2022,feng2021power}. RIS-assisted radar systems exhibit superior target detection capabilities compared to conventional radar systems without RIS. This improvement is particularly notable as the signal-to-noise ratio (SNR) is enhanced at the receiver \cite{aubry_reconfigurable_2021}, a benefit that becomes even more pronounced in non-line-of-sight (NLoS) scenarios where obstacles are present between the radar and the target \cite{song2022intelligent,10605963}. RIS-assisted radar systems can operate effectively in both monostatic and bistatic configurations \cite{liu2023integrated,buzzi_foundations_2022}. Extensive research has been conducted specifically on monostatic RIS-assisted radar setups \cite{song2022intelligent,aubry_reconfigurable_2021,grossi2023radar,shao2022target,buzzi_radar_2021}. In cases where a line-of-sight (LoS) path exists between the monostatic radar and the target, the strategic implementation of an RIS can create an additional, controlled path between the radar and the target. This is achieved by carefully designing the phases of the RIS elements, thereby boosting the SNR \cite{buzzi_radar_2021}. The integration of RISs into radar systems not only enhances their detection performance but also paves the way for new explorations in the realm of Integrated Sensing and Communications (ISAC) \cite{liu2023integrated,chepuri2023integrated}, as the digital signal processing characteristics can be shared by radar and communication systems, supporting their cooperative design \cite{chafii2023twelve}. 

Radars operate at distinct frequencies determined by the specific requirements of the surveillance mission. Lower frequencies are favored for long-range radar applications due to higher available power, and these frequencies are less affected by atmospheric attenuation. Conversely, applications that necessitate short-range surveillance benefit from higher frequencies. This preference is attributed to the capacity of antennas at higher frequencies to achieve narrower beamwidths and the suitability for operations at lower power levels \cite{salman2023tutorial}. Moreover, three-dimensional radars can be used to direct the signal accurately towards the RISs \cite{liaquat2024framework}, thereby improving the performance of the RIS-assisted radar system. 
Computing the SNR is paramount in any radar system, where it is determined for a given probability of a false alarm to achieve the desired probability of detection \cite{skolnik1962introduction}. The requisite SNR to guarantee target detection with a specific probability, considering a pre-selected false alarm probability, is plotted as receiver operating characteristic (ROC) curves, a pivotal diagnostic tool for radar systems \cite{salman2024drone}.

In \cite{10605963}, the authors derived a comprehensive equation to encapsulate the concept of free-space path loss (PL) within the scope of single RIS-assisted radars, using the principles of electromagnetic theory. 

\subsection{Main Contributions}
In this paper, we establish a theoretical framework for conducting performance analysis on radar systems supported by dual RISs. We derive the equation that quantifies the received power, SNR, and path loss in a radar system with a dual RIS. The correlation between the power received at the target and different system parameters, including the radar's transmit power, the gains of the transmitting/receiving antennas, the number of elements in the RIS, the size of each RIS element and its gain, the operating frequency, and the deployment of RISs, is discussed in detail. The contributions of this paper are presented as follows: 
\begin{itemize}
    \item We derive a mathematical expression for the received power, SNR and path loss associated with a dual-RIS-assisted radar system model from the electromagnetic (EM) theory perspective, considering the physical dimensions of the RISs, their location and the radiation patterns of their unit cells. We derive a mathematical expression that provides a closed-form solution for the SNR and path loss associated with the multiple reflections in dual RIS-assisted radar systems at L-band. In these systems, the radar employs a collection of strategically positioned RISs to detect the target by utilizing multiple reflections while all other paths are fully obstructed.

    \item We compare the SNR and path loss of dual-RIS-assisted radar and then present the results at L-band. We discuss the impact of RISs with varying unit cells in RISs compared with an equal number of unit cells on the SNR of the received signal.

\end{itemize}

The remaining sections are organized as follows. 
Section II describes the system model, including details on the setup and the derivation of expressions of received power for a dual RIS-assisted radar system. Section III presents the results and discusses their implications. Finally, Section IV concludes the paper.

\section{System Model}

\begin{figure}
\includegraphics[width=\textwidth]{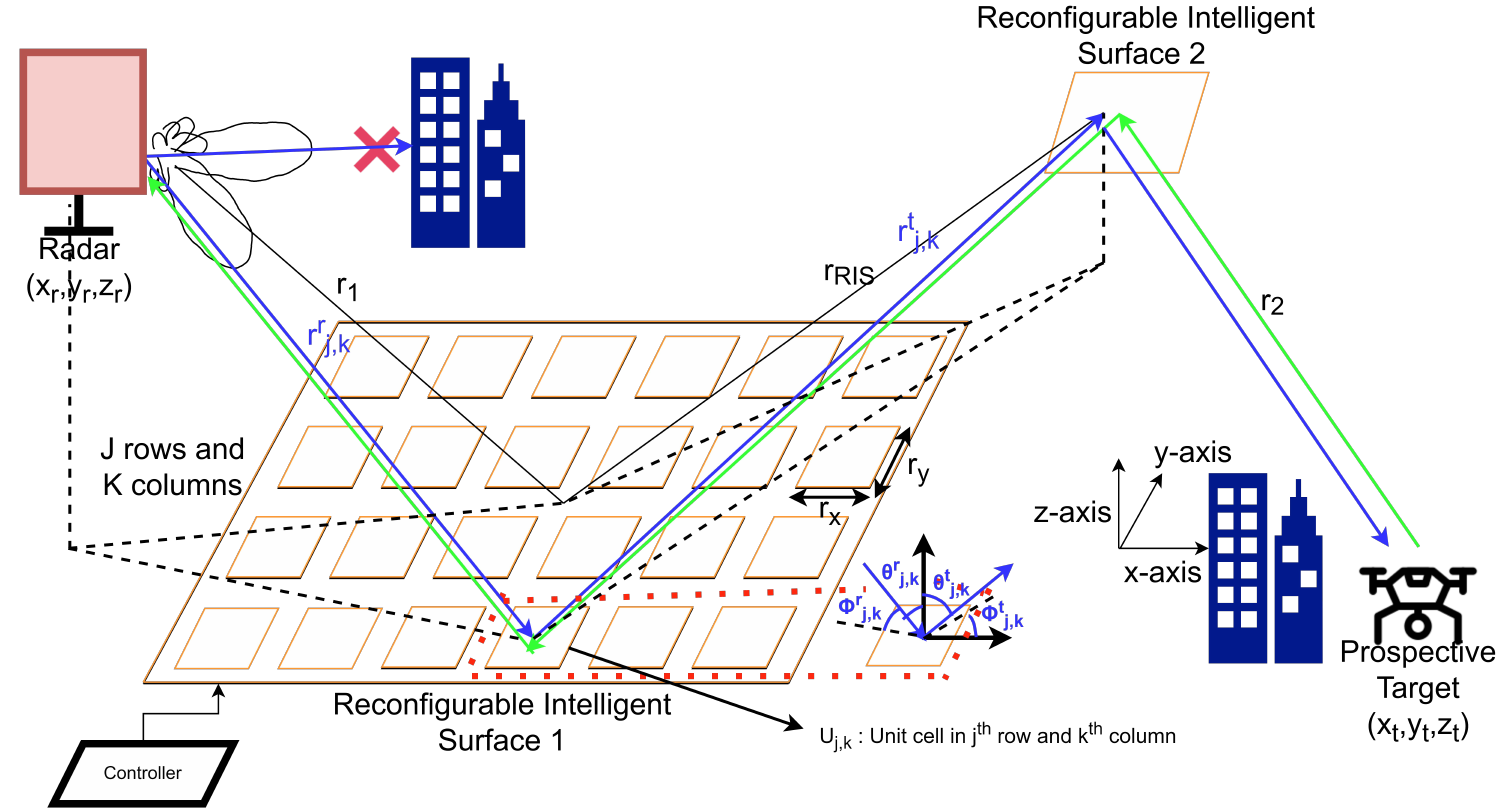}
\caption{Target detection using a dual RIS-assisted monostatic 3D radar system in a non-line-of-sight (NLoS) scenario.}
\label{fig:NLOS2}
\end{figure}

For a single RIS-assisted monostatic radar system \cite{10605963},  where no LoS path exists between the radar and the target, the RIS acts as a relay between the radar and the target, reflecting and focusing the radar signal toward the target. This configuration follows the Radar-RIS-Target-RIS-Radar path. We explore the effect of adding another RIS in the system to derive the impact of each RIS and the subsequent two hops on the SNR expression for dual RIS-assisted radar systems in an NLoS scenario as the signal follows the Radar-RIS 1-RIS 2-Target-RIS 2-RIS 1-Radar path. The system model representing a dual RIS-assisted radar system is illustrated in Fig.~\ref{fig:NLOS2}. 
Each unit cell within the RIS-1 has an x-axis dimension of $r_{x_1}$ and a y-axis dimension of $r_{y_1}$, with the spacing between the unit cells and the size of each unit cell varying between $\lambda/10$ and $\lambda/2$. The normalized power radiation pattern of the RIS's unit cell is represented by $F(\theta,\phi)$, demonstrating the dependency of the unit cell's incident / reflected power density on the incident/reflected angle. The gain of the unit cell is denoted by $G$, which is associated only with the unit cell's normalized power radiation pattern $F(\theta,\phi)$. The reflection coefficient $\Gamma_{j,k}$ of the unit cell in the $j^{th}$ row and $k^{th}$ column, represented by $U_{j,k}$, is controllable and can be expressed as,

\begin{equation} 
\small
\Gamma_{j,k}= \eta_{j,k} e^{j \phi_{j,k}} ,   
\end{equation}
\normalsize

where the amplitude and phase shift of the unit cell $U_{j,k}$ are represented by $\eta_{j,k}$ and $\phi_{j,k}$, respectively. The center position of the unit cell $U_{j,k}$ is given by $((j-\frac{1}{2})r_x,(k-\frac{1}{2})r_y,0)$. The radar is located at $(x_r,y_r,z_r)$, and the target is placed at $(x_t,y_t,z_t)$. 
It is important to observe that the distance separating the two RISs is denoted by $r_\text{RIS}$. We will determine this system's received power, SNR, and path loss using the dual RIS-assisted radar system. We introduce RIS-2 into the architecture (with MxN elements along the x-axis and y-axis, respectively), which has the gain $G_2$ and reflection coefficient $\Gamma_2$. The gain and reflection coefficient of RIS-1 is denoted as $G_1$ and $\Gamma_1$, respectively, with JxK elements along the x-axis and y-axis, respectively. Let $U_{m,n}$ represent the unit cell of RIS-2 with dimensions $r_{x_2}$ and $r_{y_2}$ along the x-axis and y-axis, respectively, and the following symbols are used to define the variables.

\begin{itemize}
    \item $(r_{x1}, r_{y1})$ – Dimensions of unit cell $U_{j,k}$ in RIS-1 along x- and y-axis
    \item $(r_{x2}, r_{y2})$ – Dimensions of unit cell $U_{m,n}$ in RIS-2 along x- and y-axis
    \item $r_1$, $r_2$ – Distances from radar to RIS-1 centre, and from RIS-2 centre to target, respectively
    \item $r^\text{RIS}$ – Distance between centres of RIS-1 and RIS-2
    \item $(\theta_r, \phi_r)$ – Elevation and azimuth angles from RIS-1 centre to radar
    \item $(\theta_t, \phi_t)$ – Elevation and azimuth angles from RIS-2 centre to target
    \item $r^r_{j,k}$ – Distance between unit cell $U_{j,k}$ (RIS-1) and radar
    \item $r^\text{RIS}_{j,k}$, $r^\text{RIS}_{m,n}$ – Distances between unit cells $U_{j,k}$ (RIS-1) and $U_{m,n}$ (RIS-2)
    \item $r^t_{m,n}$ – Distance between unit cell $U_{m,n}$ (RIS-2) and target
    \item $(\theta^{r}_{1,j,k}, \phi^{r}_{1,j,k})$ – Elevation and azimuth angles from $U_{j,k}$ (RIS-1) to radar
    \item $(\theta^{r}_{2,j,k}, \phi^{r}_{2,j,k})$ – Elevation and azimuth angles from $U_{j,k}$ (RIS-1) to $U_{m,n}$ (RIS-2)
    \item $(\theta^t_{1,m,n}, \phi^t_{1,m,n})$ – Elevation and azimuth angles from $U_{m,n}$ (RIS-2) to $U_{j,k}$ (RIS-1)
    \item $(\theta^t_{2,m,n}, \phi^t_{2,m,n})$ – Elevation and azimuth angles from $U_{m,n}$ (RIS-2) to target
    \item $(\theta_\text{RIS}, \phi_\text{RIS})$ – Elevation and azimuth angles between RIS-1 and RIS-2 centres
\end{itemize}

It is to be noted that for far-field approximation, $r^\text{RIS}_{j,k}$ = $r^\text{RIS}_{m,n}$, $\theta^{r}_{2,j,k}$ = $\theta^t_{1,m,n}$, and $\phi^{r}_{2,j,k}$ = $\phi^t_{1,m,n}$, but we have mentioned them separately for a better understanding of the derivation of power expression. We now derive the power received at the radar receiver step by step, tracing the Transmitter-RIS 1-RIS 2-Target-RIS 2-RIS 1-Receiver path.

As illustrated in Fig.~\ref{fig:NLOS2}, the radar transmitter emits a signal of wavelength $\lambda$ with power $P_t$ through an antenna with a normalized power radiation pattern of $F_{\text{rad}}(\theta,\phi)$ and gain $G_t$. The signal is transmitted to the RIS-1, which reflects it towards RIS-2 which further reflects it towards the target. A portion of the signal is then reflected towards the RIS-2 by the target, dependent on the target's RCS denoted by $\sigma$. The received signal is then reflected back to the RIS-2 and to the radar receiver antenna by the RIS-2, with a normalized power radiation pattern of $F_{\text{rad}}(\theta,\phi)$ and gain $G_t$, as the radar exhibits reciprocity, i.e., the radiation pattern is identical during transmission and reception and the same antenna is being used for both transmission and reception in a monostatic radar. Therefore, the signal hops twice from the same RIS during the target detection in our framework.

We calculate the power transfer between the radar and the RIS-1 unit cell, between the RIS-1 unit cell and the RIS-2 unit cell, across the entire RIS-1 array to the RIS-2 unit cell, from the RIS-2 unit cell to the target, between the complete RIS-2 array and the target, from the target to the RIS-2 unit cell, between the RIS-2 unit cell and the RIS-1 unit cell, across the entire RIS-2 array to the RIS-1 unit cell, and finally from RIS-1 to the radar receiver, as derived in the following analysis.

The power received from the radar transmitter at a single unit cell of RIS-1, denoted as $P_{1}$, is given by
\begin{equation}
P_{1} = \frac{P_t G_t F_{\text{rad}}(\theta^{r}_{1,j,k},\phi^{r}_{1,j,k}) F(\theta^{r}_{1,j,k},\phi^{r}_{1,j,k}) r_{x_1} r_{y_1}}{4\pi {r^r_{j,k}}^2}.
\end{equation}

The power received from a RIS-1 unit cell at a single unit cell of RIS-2, considering the propagation distance $r_\text{RIS}$, is expressed as
\begin{align}
P_{2} = &\frac{P_t G_t F_{\text{rad}}(\theta^{r}_{1,j,k},\phi^{r}_{1,j,k}) F(\theta^{r}_{1,j,k},\phi^{r}_{1,j,k}) r_{x_1} r_{y_1}}{(4\pi)^2 {r^r_{j,k}}^2 {r^\text{RIS}_{j,k}}^2} \nonumber \\
& \times |\Gamma^2_{1,j,k}| G_1 F(\theta^{r}_{2,j,k},\phi^{r}_{2,j,k}) F(\theta^{t}_{1,m,n},\phi^{t}_{1,m,n}) r_{x_2} r_{y_2}.
\end{align}

The power received from the complete RIS-1 array at a single unit cell of RIS-2, denoted as $P_{3}$, can be written as
\begin{multline}
P_3 =  \frac{P_t {G_t} G_1 {r_{x_1}} {r_{y_1}} {r_{x_2}} {r_{y_2}} F(\theta^{t}_{1,m,n},\phi^{t}_{1,m,n})}{(4\pi)^2}
 \left|\sum_{j=1}^{J}\sum_{k=1}^{K} \frac{\sqrt{F^{\text{dual RIS}}_{\text{combine},j,k}} \eta_{1,j,k}}{r^r_{j,k} r^\text{RIS}_{j,k}} e^{-j\left(\frac{ 2 \pi r^r_{j,k} }{\lambda}-\phi_{1,j,k}+\frac{ 2 \pi r^\text{RIS}_{j,k} }{\lambda}\right)}\right|^2,
\end{multline}
where
\begin{equation}
F^{\text{dual RIS}}_{\text{combine},j,k} = F_{\text{rad}}(\theta^{r}_{1,j,k},\phi^{r}_{1,j,k}) F(\theta^{r}_{1,j,k},\phi^{r}_{1,j,k}) F(\theta^{r}_{2,j,k},\phi^{r}_{2,j,k}).
\end{equation}

The power directed by the RIS-2 unit cell towards the target, denoted as $P_4$, based on $P_3$, is given by
\begin{equation}
P_4 = P_3 |\Gamma^2_{2,m,n}| G_2 \frac{F(\theta^{t}_{2,m,n},\phi^{t}_{2,m,n})}{4\pi{r^t_{n,m}}^2}.
\end{equation}

The power received at the target from the complete RIS-2 array, denoted as $P_{5}$, is
\begin{multline}
P_5 =  \frac{P_t {G_t} G_1 G_2 {r_{x_1}} {r_{y_1}} {r_{x_2}} {r_{y_2}}}{(4\pi)^3}  \left|\sum_{j=1}^{J}\sum_{k=1}^{K} \frac{\sqrt{F^{\text{dual RIS}}_{\text{combine},j,k}} \eta_{1,j,k}}{r^r_{j,k} r^\text{RIS}_{j,k}} e^{-j\left(\frac{ 2 \pi r^r_{j,k} }{\lambda}-\phi_{1,j,k}+\frac{ 2 \pi r^\text{RIS}_{j,k} }{\lambda}\right)}\right|^2 \\
\times \left|\sum_{m=1}^{M}\sum_{n=1}^{N} \frac{\sqrt{F_{\text{combine},m,n}} \eta_{2,m,n}}{r^t_{m,n}} e^{-j\left(\frac{ 2 \pi r^t_{m,n} }{\lambda}-\phi_{2,m,n}\right)}\right|^2,
\end{multline}
with
\begin{equation}
F_{\text{combine},m,n} = F(\theta^{t}_{1,m,n},\phi^{t}_{1,m,n}) F(\theta^{t}_{2,m,n},\phi^{t}_{2,m,n}).
\end{equation}

The power received from a RIS-2 unit cell at a RIS-1 unit cell, denoted as $P_{7}$, is given by
\begin{align}
P_{6} = &\frac{ P_5 \sigma}{4 \pi {r^t_{m,n}}^2 } F(\theta^{t}_{2,m,n},\phi^{t}_{2,m,n}) {r_{x_2}}{r_{y_2}} F(\theta^{r}_{2,j,k},\phi^{r}_{2,j,k}) r_{x_1} r_{y_1}   |\Gamma^2_{2,m,n}| \frac{G_2}{4\pi {r^\text{RIS}_{m,n}}^2 } F(\theta^{t}_{1,m,n},\phi^{t}_{1,m,n}).
\end{align}

The total power received from the complete RIS-2 array back at a RIS-1 unit cell, $P_{7}$, can be written as
\begin{multline}
P_{7} =  \frac{P_t {G_t} \sigma G_1 {G_2}^2 {r_{x_1}^2} {r_{y_1}^2} {r_{x_2}^2} {r_{y_2}^2}}{(4\pi)^5} F(\theta^{r}_{2,j,k},\phi^{r}_{2,j,k})  \left|\sum_{j=1}^{J}\sum_{k=1}^{K} \frac{\sqrt{F^{\text{dual RIS}}_{\text{combine},j,k}} \eta_{1,j,k}}{r^r_{j,k} r^\text{RIS}_{j,k}} e^{-j\left(\frac{ 2 \pi r^r_{j,k} }{\lambda}-\phi_{1,j,k}+\frac{ 2 \pi r^\text{RIS}_{j,k} }{\lambda}\right)}\right|^2 \\
\times \left|\sum_{m=1}^{M}\sum_{n=1}^{N} \frac{\sqrt{F_{\text{combine},m,n}} \eta_{2,m,n}}{r^t_{m,n} \sqrt{r^\text{RIS}_{m,n}}} e^{-j\left(\frac{ 4 \pi r^t_{m,n} }{\lambda}-\phi_{2,m,n} + \frac{ 2 \pi r^\text{RIS}_{m,n} }{\lambda} - \phi^{\prime}_{2,m,n}\right)}\right|^4.
\end{multline}

Defining
\begin{equation}
W_{j,k} = \frac{\sqrt{F^{\text{dual RIS}}_{\text{combine},j,k}} \eta_{1,j,k}}{r^r_{j,k} \sqrt{r^\text{RIS}_{j,k}}} e^{-j\left(\frac{ 4 \pi r^r_{j,k} }{\lambda}-\phi_{1,j,k}+\frac{ 2 \pi r^\text{RIS}_{j,k} }{\lambda}-\phi^{\prime}_{1,j,k}\right)},
\end{equation}
and
\begin{equation}
W_{m,n} = \frac{\sqrt{F_{\text{combine},m,n}} \eta_{2,m,n}}{r^t_{m,n} \sqrt{r^\text{RIS}_{m,n}}} e^{-j\left(\frac{ 4 \pi r^t_{m,n} }{\lambda}-\phi_{2,m,n}+\frac{ 2 \pi r^\text{RIS}_{m,n} }{\lambda}-\phi^{\prime}_{2,m,n}\right)},
\end{equation}
the power received from the complete RIS-1 array at the radar is
\begin{multline}
P^{\text{dual RIS}}_\text{received} = \frac{P_t \sigma \lambda^2 {G_t}^2 {G_1}^2 {G_2}^2 {r_{x_1}^2} {r_{y_1}^2} {r_{x_2}^2} {r_{y_2}^2}}{(4\pi)^7}  \left|\sum_{j=1}^{J}\sum_{k=1}^{K} W_{j,k}\right|^4 \left|\sum_{m=1}^{M}\sum_{n=1}^{N} W_{m,n}\right|^4.
\end{multline}

Considering the known angles between radar and RIS-1, and RIS-1 and RIS-2, the maximum received power is achieved for $\theta_t = \theta_\text{RIS}$ and $\phi_t = \phi_\text{RIS} + \pi$, leading to
\begin{multline}
P^{\text{dual RIS}}_{\text{received, max}} = \left[\frac{P_t G_t^2 \lambda^2 \sigma}{(4 \pi)^3 {r_1}^4}\right] \left[\frac{G_1^2 r_{x1}^2 r_{y1}^2 J^4 K^4 \eta_1^4 {F^{\text{dual RIS}}_{\text{combine},j,k}}^2}{(4 \pi)^2 r_{2}^4}\right]  \left[\frac{G_2^2 r_{x_2}^2 r_{y_2}^2 M^4 N^4 \eta_2^4 {F_{\text{combine},m,n}}^2}{(4 \pi)^2 r_\text{RIS}^4}\right].
\end{multline}

Finally, the SNR for the dual RIS-assisted radar system integrating $P_N$ pulses is
\begin{align}
\text{SNR}_{\text{dual RIS}} &= \left[\frac{P_t G_t^2 \lambda^2 \sigma {P_N}}{(4 \pi)^3  k T_0 B L}\right] \left[\frac{G_1^2 r_{x1}^2 r_{y1}^2}{(4 \pi)^2}\right] \left[\frac{G_2^2 r_{x_2}^2 r_{y_2}^2}{(4 \pi)^2}\right] \nonumber \left|\sum_{j=1}^{J}\sum_{k=1}^{K} W_{j,k}\right|^4 \left|\sum_{m=1}^{M}\sum_{n=1}^{N} W_{m,n}\right|^4,
\end{align}
which under maximum alignment becomes
\begin{multline}
\text{SNR}_{\text{dual RIS},\text{max}} = \left[\frac{P_t G_t^2 \lambda^2 \sigma {P_N}}{(4 \pi)^3 {r_1}^4 k T_0 B L}\right] \left[\frac{G_1^2 r_{x1}^2 r_{y1}^2 J^4 K^4 \eta_1^4 {F^{\text{dual RIS}}_{\text{combine},j,k}}^2}{(4 \pi)^2 r_{2}^4}\right] \left[\frac{G_2^2 r_{x_2}^2 r_{y_2}^2 M^4 N^4 \eta_2^4 {F_{\text{combine},m,n}}^2}{(4 \pi)^2 r_\text{RIS}^4}\right].
\end{multline}

\begin{table}
\caption{Parameters used for dual RIS-assisted radar system simulation.}
\label{table1}
\begin{tabular}{|p{5cm}|p{6cm}|}
\hline
\textbf{Parameter} & \textbf{Value} \\
\hline
Radar frequency ($f$) & L-band \\
\hline
Transmitted power ($P_t$) & 30 dBW \\
\hline
Antenna gain ($G_t$) & 30 dB \\
\hline
Number of pulses for coherent integration ($P_\text{N}$) & 1, 40, 80 \\
\hline
RIS gains ($G_1$, $G_2$) & 4 dB \\
\hline
Inter unit cell spacing ($r_{x1}, r_{y1}, r_{x2}, r_{y2}$) & $\lambda/2$ \\
\hline
Target RCS ($\sigma$) & 0.02 m\textsuperscript{2} \\
\hline
Unit cells in RIS-1 (J, K) & 10, 19, 28, 37, 46\\
\hline
Unit cells in RIS-2 (M, N) & 10, 19, 28, 37, 46\\
\hline
Unit cell amplitude ($\eta_1, \eta_2$) & 0.8 \\
\hline
Distance between radar and RIS-1 ($r_1$) & 250 m, 1750 m \\
\hline
Distance between RIS-1 and RIS-2 ($r_\text{RIS}$) & 50 m \\
\hline
\end{tabular}
\end{table}

Therefore, the SNR is directly proportional to the transmitted power, the square of the antenna gains, the radar cross-section, and the square of the wavelength. Conversely, the SNR is inversely proportional to the noise figure of the radar receiver and RIS, the losses in the radar system and path, temperature, and bandwidth, and the fourth power of the distances between the radar and RIS, amongst the RISs and RIS and the target. This distance is the main reason for power reduction as it is the fourth power to the products of these distances.

\vspace{-0.3cm}
\section{Simulation Results and Discussion}

The performance evaluation of the derived equations for SNR can be comprehensively assessed through simulation, focusing on the detection of a drone target with a radar cross-section ($\sigma$) of 0.02 m².  The simulation parameters are presented in Table \ref{table1}. For the simulation, the radiation pattern of the RIS is modelled with a gain of 4 dB and a half-power beamwidth of 45 degrees. At L-band, an inter-unit cell spacing of $\lambda/2$ (approximately 0.107 m) is considered, resulting in RIS configurations with varying numbers of unit cells: 10, 19, 28, 37, and 46. These configurations correspond to RIS dimensions of approximately 1m x 1m, 2m x 2m, 3m x 3m, 4m x 4m, and 5m x 5m, respectively.

\begin{figure}
\includegraphics[width=\textwidth]{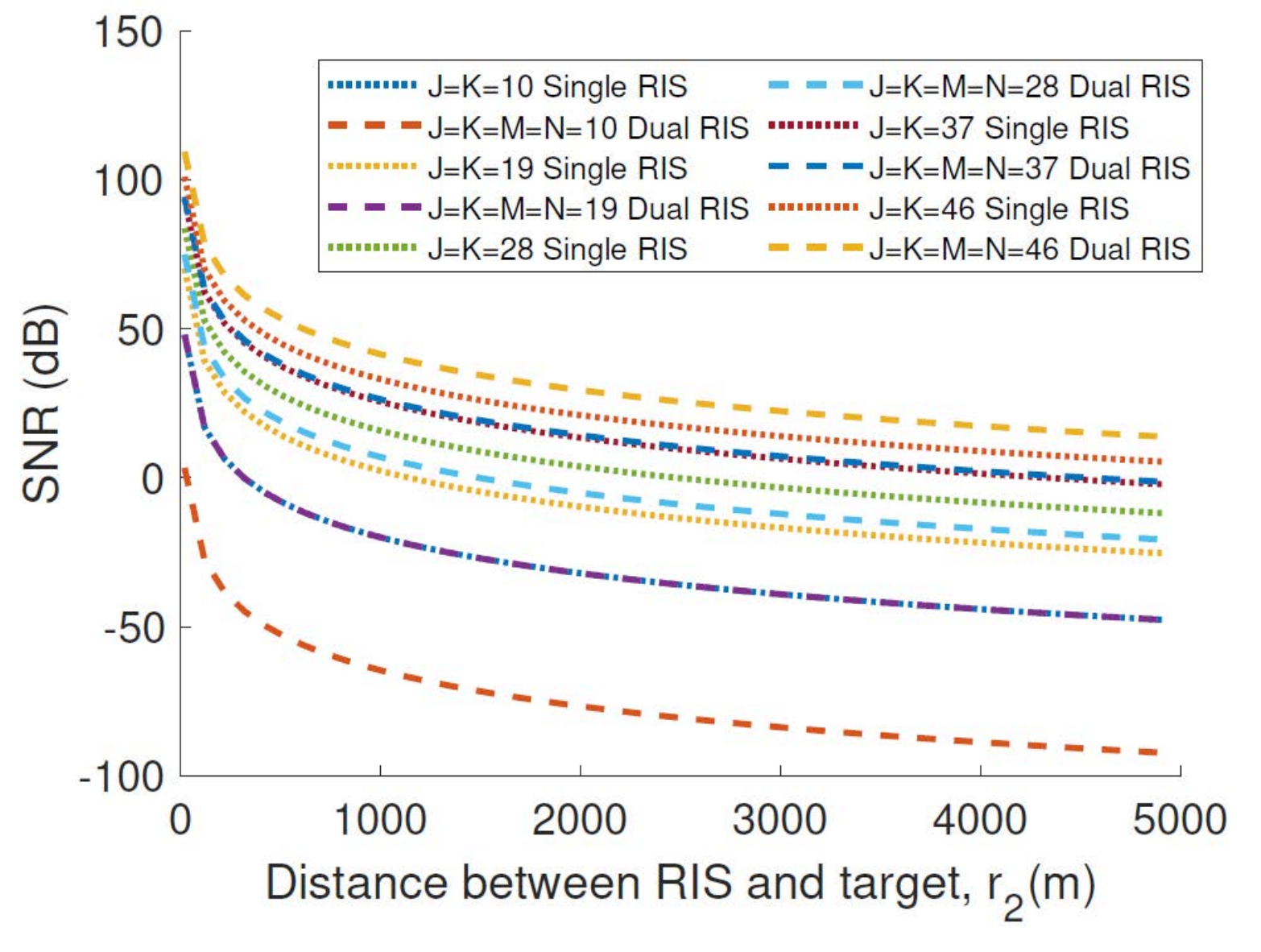}
\caption{SNR performance comparison between single and dual RIS-assisted radar systems.}
\label{fig:fig4}
\vspace{-1cm}
\end{figure}

To compare the performance of single and dual RIS-assisted radars, Fig.~\ref{fig:fig4} displays the SNR as the distance between the RIS and the target increases. As expected, the SNR decreases with an increasing distance between the target and the RIS. Moreover, the SNR improves as the number of elements in the RIS increases, enhancing the overall performance of the RIS-assisted radar system. The vertical lines in Fig.~\ref{fig:fig4} indicate the minimum distance required for the far-field assumption to hold, which is a crucial parameter for determining radar system accuracy. Comparing the performance of single and dual RIS-assisted radar systems, it is evident that a perfectly aligned dual RIS-assisted system outperforms its single RIS-assisted counterpart as the number of elements increases and is the RIS and the radar are placed at specific locations and the target angle is such that it is reflected to the RIS at the maximum value. Specifically, for RIS configurations with 37x37 and 46x46 elements at L-band, the SNR is superior for the dual RIS system. This improvement can be attributed to the enhanced RIS effect, which increases with a higher number of elements and surpasses unity for certain RIS sizes, as demonstrated for L-band in Table \ref{table2}. Therefore, for each subsequent RIS addition, with the RISs positioned in a way such that their normalized radiation patterns completely overlap, the addition of each RIS enhances the system performance.

It is important to note that while dual RIS-assisted radar systems theoretically provide higher SNR due to increased reflection gain, this advantage is not universal. As observed in Fig.~\ref{fig:fig4}, for smaller RIS sizes such as $10 \times 10$ and $19 \times 19$, the SNR achieved using a dual RIS configuration is lower than that of the single RIS configuration. This is attributed to the increased path loss associated with the dual RIS setup, which involves multiple signal propagation stages. Specifically, the signal traverses a longer path—Radar $\rightarrow$ RIS-1 $\rightarrow$ RIS-2 $\rightarrow$ Target $\rightarrow$ RIS-2 $\rightarrow$ RIS-1 $\rightarrow$ Radar—resulting in a higher cumulative free-space attenuation. For smaller RIS sizes, the array gain is insufficient to compensate for this additional loss. However, as the RIS dimensions increase (e.g., $37 \times 37$ and $46 \times 46$), the dual RIS system provides superior SNR performance due to enhanced beamforming gain across both RIS surfaces.

\begin{table}
\caption{RIS configuration, size, and effect on SNR.}
\label{table2}
\begin{tabular}{|p{2cm}|p{2cm}|p{3cm}|p{2cm}|p{2cm}|}
\hline
\textbf{RIS configuration} & \textbf{Number of elements} & \textbf{Approximate RIS size at L-band} & \textbf{Far-field distance (m)} & \textbf{RIS effect (dB)} \\
\hline
1 & 10 $\times$ 10 & 1 m $\times$ 1 m & 10.7142 & -44.62 \\
\hline
2 & 19 $\times$ 19 & 2 m $\times$ 2 m & 38.6631 & -22.32 \\
\hline
3 & 28 $\times$ 28 & 3 m $\times$ 3 m & 83.9664 & -8.85 \\
\hline
4 & 37 $\times$ 37 & 4 m $\times$ 4 m & 146.6199 & 0.84 \\
\hline
5 & 46 $\times$ 46 & 5 m $\times$ 5 m & 226.6236 & 8.4 \\
\hline
\end{tabular}
\end{table}

It is critical to note that in RIS-assisted radar systems, the overall path loss for the radar-RIS-target-RIS-radar path is not simply the sum of individual path losses. Instead, it results from multiplying the path losses across the radar-RIS and RIS-target segments twice, as the signal has to traverse the path twice to return to the radar. To effectively reduce the impact of path loss caused by multiplicative fading, employing a considerably expanded RIS of many unit cells is imperative to attain a desirable level of array gain.

\section{Conclusion}

Using dual RIS can enhance the ability to detect targets for an existing radar system. By carefully tuning the phase and amplitude of the signals reflected by the RISs, it is possible to create a constructive interference pattern that boosts the received signal strength. RIS-assisted radar systems have the potential to significantly improve radar performance by enhancing signal power at the receiver. In various scenarios, strategically placed RISs can extend radar coverage by facilitating the detection of targets behind obstacles and tracking moving targets such as aircraft and vehicles. In certain scenarios, using two or more RISs in conjunction with radar systems can yield superior performance compared to a single RIS-assisted radar. This improvement is particularly beneficial in scenarios where conventional radar systems or single RIS-assisted radars struggle to detect targets accurately due to the absence of a direct path to the target. By strategically coordinating the deployment of multiple RISs, radar coverage can be expanded, offering substantial advantages over traditional radar systems. The successful integration of multiple RISs with radar systems holds promise for advancing radar technology in diverse real-world scenarios. As RIS technology evolves, many potential applications and use cases may emerge, presenting new opportunities to optimize radar system performance in various environments. 

\textbf{Acknowledgements} The authors would like to thank Ministry of Higher Education Malaysia for Fundamental Research Grant Scheme with Project Code FRGS/1/2022/TK07/USM/02/14 for permitting them to carry out this research.

\bibliographystyle{splncs03_unsrt.bst}
% \bibliography{MyLibrary}

\begin{thebibliography}{10}
\providecommand{\url}[1]{\texttt{#1}}
\providecommand{\urlprefix}{URL }

\bibitem{ElMossallamy_reconfigurable_2020}
ElMossallamy, M.A., Zhang, H., Song, L., Seddik, K.G., Han, Z., Li, G.Y.: Reconfigurable intelligent surfaces for wireless communications: Principles, challenges, and opportunities. IEEE Transactions on Cognitive Communications and Networking  6(3),  990--1002 (2020)

\bibitem{tang2022path}
Tang, W., Chen, X., Chen, M.Z., Dai, J.Y., Han, Y., Di~Renzo, M., Jin, S., Cheng, Q., Cui, T.J.: Path loss modeling and measurements for reconfigurable intelligent surfaces in the millimeter-wave frequency band. IEEE Transactions on Communications  70(9),  6259--6276 (2022)

\bibitem{liu_reconfigurable_intelligent_2021}
Liu, Y., Liu, X., Mu, X., Hou, T., Xu, J., Di~Renzo, M., Al-Dhahir, N.: Reconfigurable intelligent surfaces: Principles and opportunities. IEEE Communications Surveys \& Tutorials  23(3),  1546--1577 (2021)

\bibitem{bjornson_reconfigurable_2020}
Bj\"ornson, E., \"Ozdogan, O., Larsson, E.G.: Reconfigurable intelligent surfaces: Three myths and two critical questions. IEEE Communications Magazine  58(12),  90--96 (2020)

\bibitem{basar2019wireless}
Basar, E., Di~Renzo, M., De~Rosny, J., Debbah, M., Alouini, M.S., Zhang, R.: Wireless communications through reconfigurable intelligent surfaces. IEEE Access  7,  116753--116773 (2019)

\bibitem{zhang2020optically}
Zhang, X.G., Jiang, W.X., Jiang, H.L., Wang, Q., Tian, H.W., Bai, L., Luo, Z.J., Sun, S., Luo, Y., Qiu, C.W., et~al.: An optically driven digital metasurface for programming electromagnetic functions. Nature Electronics  3(3),  165--171 (2020)

\bibitem{wu_towards_2020}
Wu, Q., Zhang, R.: Towards smart and reconfigurable environment: Intelligent reflecting surface aided wireless network. IEEE Communications Magazine  58(1),  106--112 (2020)

\bibitem{di_renzo_2020_smart_radio}
Di~Renzo, M., Zappone, A., Debbah, M., Alouini, M.S., Yuen, C., de~Rosny, J., Tretyakov, S.: Smart radio environments empowered by reconfigurable intelligent surfaces: How it works, state of research, and the road ahead. IEEE Journal on Selected Areas in Communications  38(11),  2450--2525 (2020)

\bibitem{buzzi_radar_2021}
Buzzi, S., Grossi, E., Lops, M., Venturino, L.: Radar target detection aided by reconfigurable intelligent surfaces. IEEE Signal Processing Letters  28,  1315--1319 (2021)

\bibitem{di_renzo_communication_2022}
Di~Renzo, M., Danufane, F.H., Tretyakov, S.: Communication models for reconfigurable intelligent surfaces: From surface electromagnetics to wireless networks optimization. Proceedings of the IEEE  110(9),  1164--1209 (2022)

\bibitem{lu_target_2021}
Lu, W., Lin, Q., Song, N., Fang, Q., Hua, X., Deng, B.: Target detection in intelligent reflecting surface aided distributed mimo radar systems. IEEE Sensors Letters  5(3),  1--4 (2021)

\bibitem{buzzi_foundations_2022}
Buzzi, S., Grossi, E., Lops, M., Venturino, L.: Foundations of mimo radar detection aided by reconfigurable intelligent surfaces. IEEE Transactions on Signal Processing  70,  1749--1763 (2022)

\bibitem{haobo_metardar_2022}
Zhang, H., Zhang, H., Di, B., Bian, K., Han, Z., Song, L.: Metaradar: Multi-target detection for reconfigurable intelligent surface aided radar systems. IEEE Transactions on Wireless Communications  21(9),  6994--7010 (2022)

\bibitem{feng2021power}
Feng, Z., Wang, B., Zhao, Y., Luan, M., Hu, F.: Power optimization for target localization with reconfigurable intelligent surfaces. Signal Processing  189,  108252 (2021)

\bibitem{aubry_reconfigurable_2021}
Aubry, A., De~Maio, A., Rosamilia, M.: Reconfigurable intelligent surfaces for n-los radar surveillance. IEEE Transactions on Vehicular Technology  70(10),  10735--10749 (2021)

\bibitem{song2022intelligent}
Song, X., Xu, J., Liu, F., Han, T.X., Eldar, Y.C.: Intelligent reflecting surface enabled sensing: Cram$\backslash$'er-rao bound optimization. arXiv preprint arXiv:2207.05611  (2022)

\bibitem{10605963}
Liaquat, S., Naqvi, I.H., Mahyuddin, N.M., Hassan, N.U.: The impact of swerling models on snr and path loss in ris-assisted monostatic radar under nlos conditions. In: 2024 47th International Conference on Telecommunications and Signal Processing (TSP). pp. 69--74 (2024)

\bibitem{liu2023integrated}
Liu, R., Li, M., Luo, H., Liu, Q., Swindlehurst, A.L.: Integrated sensing and communication with reconfigurable intelligent surfaces: Opportunities, applications, and future directions. IEEE Wireless Communications  30(1),  50--57 (2023)

\bibitem{grossi2023radar}
Grossi, E., Taremizadeh, H., Venturino, L.: Radar target detection and localization aided by an active reconfigurable intelligent surface. IEEE Signal Processing Letters  (2023)

\bibitem{shao2022target}
Shao, X., You, C., Ma, W., Chen, X., Zhang, R.: Target sensing with intelligent reflecting surface: Architecture and performance. IEEE Journal on Selected Areas in Communications  40(7),  2070--2084 (2022)

\bibitem{chepuri2023integrated}
Chepuri, S.P., Shlezinger, N., Liu, F., Alexandropoulos, G.C., Buzzi, S., Eldar, Y.C.: Integrated sensing and communications with reconfigurable intelligent surfaces: From signal modeling to processing. IEEE Signal Processing Magazine  40(6),  41--62 (2023)

\bibitem{chafii2023twelve}
Chafii, M., Bariah, L., Muhaidat, S., Debbah, M.: Twelve scientific challenges for 6g: Rethinking the foundations of communications theory. IEEE Communications Surveys \& Tutorials  (2023)

\bibitem{salman2023tutorial}
Liaquat, S., Mahyuddin, N.M., Naqvi, I.H.: An end-to-end modular framework for radar signal processing: A simulation-based tutorial. IEEE Aerospace and Electronic Systems Magazine pp. 1--17 (2023)

\bibitem{liaquat2024framework}
Liaquat, S., Faizan, M., Chattha, J.N., Butt, F.A., Mahyuddin, N.M., Naqvi, I.H.: A framework for preventing unauthorized drone intrusions through radar detection and gps spoofing. Ain Shams Engineering Journal  15(5),  102707 (2024)

\bibitem{skolnik1962introduction}
Skolnik, M.I.: Introduction to radar. Radar handbook  2, ~21 (1962)

\bibitem{salman2024drone}
Liaquat, S., Mahyuddin, N.M., Naqvi, I.H.: Drone detection using swerling-i model with l-band/x-band radar in free space and raining scenario. In: International Conference on Robotics, Vision, Signal Processing and Power Applications. pp. 421--427. Springer (2024)

\end{thebibliography}

\vspace{0.5em}

\end{document}